\documentclass[a4paper,english]{article}
\usepackage{babel}
\usepackage[latin1]{inputenc}
\usepackage[T1]{fontenc}
\usepackage{times}
\usepackage[pdftex]{graphicx}
\usepackage{color}
\usepackage[pdftex,colorlinks=true,citecolor=black,
            pagecolor=black,linkcolor=black,menucolor=black,
            urlcolor=black]{hyperref}
\usepackage{amsfonts}
\usepackage{amsmath}
\usepackage{amsbsy}
\usepackage{eucal}
\usepackage{caption}
\usepackage{parskip}
\usepackage{subcaption}
\usepackage{longtable}
\usepackage{url}
\usepackage[noend]{algpseudocode}
\usepackage{listings}
\usepackage[titletoc]{appendix}
\usepackage{natbib}

\usepackage{tikz,pgfplots}

\usetikzlibrary{calc}

\usetikzlibrary{external}
\tikzset{external/system call={lualatex
	\tikzexternalcheckshellescape -halt-on-error -interaction=batchmode
	-jobname "\image" "\texsource"}}

\usepackage{soul}

\newcommand{\vc}[1] { \mathbf{#1} }
\newcommand{\vs}[1] { \boldsymbol{#1} }
\newcommand{\tp}{\mathsf{T}}
\newcommand{\ti}[1] { \tilde{#1} }

\newcommand{\given} { \,|\, }

\newcommand{\data} {D}

\newcommand{\mean}[2][] { \mathrm{E}_{#1} {\left[#2\right]} }

\newcommand{\KL}[2] { \mathrm{KL} {\left(#1 \, \| \, #2\right)} }

\newcommand{\Normal}[1] { \mathrm{N} {\left(#1\right)}  } 

\newcommand{\halfStudent}[2][] { t_{#1}^+ {\left(#2\right)} }

\newcommand{\halfCauchy}[1] { \mathrm{C}^+ {\left(#1\right)} }

\newcommand{\StudentNP}[1][] { t_{#1} }
\newcommand{\halfStudentNP}[1][] { t_{#1}^+ }

\title{Projection predictive variable selection using Stan+R}
\author{Juho Piironen\footnote{first.last@aalto.fi}\; and Aki Vehtari}

\begin{document}
\maketitle

\begin{abstract}
This document is additional material to our previous study comparing several strategies for variable subset selection \citep{piironen2015}.
Our recommended approach was to fit the full model with all the candidate variables and best possible prior information, and perform the variable selection using the projection predictive framework.
Here we give an example of performing such an analysis, using Stan for fitting the model, and R for the variable selection.
\end{abstract}

\section{Introduction}

Identifying relevant explanatory variables is often of interest in applied statistical analysis.
In this context, the relevance of a variable is usually determined by its predictive power on the target variable of interest.
The gain from successful variable selection is typically improved model interpretability, but depending on the problem, also reduced measurement costs and computational savings may be obtained.

In our recect study \citep{piironen2015}, we discussed and compared several strategies for performing variable selection in practical problems and concluded that often best results are obtained by the following strategy: fit the full model with all the candidate variables and best possible prior information, and perform the variable selection using the projection predictive method \citep{goutis1998,dupuis2003}.
Our comparison showed that this approach generally outperforms the selection of the most probable variables or variable combination, methods which are often used for performing Bayesian variable selection.%

This document serves as additional material to our study, the purpose being to provide an example of how to carry out the model fitting with Stan and the subsequent variable selection using R.
Sampling the full model under the popular spike-and-slab prior \citep{mitchell1988} is infeasible with Stan, but the prior assumptions about the sparsity can be conveniently formulated using the hierarchical shrinkage priors, such as the horseshoe \citep{carvalho2009,carvalho2010}, which allow convenient computation using Stan.
All the relevant codes are provided in the Appendix.

The document is organized as follows.
Section~\ref{sec:hier_shrinkage} reviews the hierarchical shrinkage in the context of linear regression using half-Student-$t$ priors for the weight scales, and discusses the horseshoe prior as a special case.
Section~\ref{sec:projection} shortly reviews the projection predictive variable selection and discusses the computations in the example case of a linear Gaussian model.
Finally, in Section~\ref{sec:numexample} we provide an illustrative numerical example.

\section{Hierarchical shrinkage}
\label{sec:hier_shrinkage}

Consider the single output linear Gaussian regression model with several input variables, given by
\begin{align}
\begin{split}
	f_i &= \vc w^\tp \vc x_i \\
	y_i &= f_i + \varepsilon_i, \quad \varepsilon \sim \Normal{0,\sigma^2}, \quad i=1,\dots,n %
\end{split}
\label{eq:lgm}
\end{align}
where $\vc x$ is the $m$-dimensional vector of inputs, $\vc w$ contains the corresponding weights and $\sigma^2$ is the noise variance.
A hierarchical shrinkage (HS) prior for the regression weights $\vc w = (w_1,\dots,w_m)$ can be obtained as
\begin{align}
\begin{split}
	w_i \given \lambda_i,\tau &\sim \Normal{0,\lambda_i^2 \tau^2} \\
	\lambda_i &\sim \halfStudent[\nu]{0,1} \,.
\end{split}
\label{eq:hs}
\end{align}
where  $\halfStudentNP[\nu]$ denotes the half-Student-$t$ prior with $\nu$ degrees of freedom.
We will refer to \eqref{eq:hs} by the acronym HS-$t_\nu$.
The horsehoe prior \citep{carvalho2009,carvalho2010} is obtained by setting $\nu=1$, that is, by introducing half-Cauchy priors for the local scale parameters $\lambda_i$.
The horseshoe prior has been shown to possess desirable theoretical properties and good performance in practice \citep{carvalho2009,carvalho2010,datta2013,varDerPas2014}.
Intuitively, we expect the local variance parameters $\lambda_i^2$ to be large for those inputs that have high relevance, and small for those with negligible relevance, while the global variance term $\tau^2$ adjusts the overall sparsity level.
The shrinkage coefficient $\kappa_i = 1/(1+\lambda_i^2)$ describes the amount of shrinkage for the weight $w_i$, so that $\kappa_i = 0$ means no shrinkage ($\lambda_i^2$ large) and $\kappa_i=1$ complete shrinkage ($\lambda_i^2$ small).

\begin{figure}%
\centering
	\includegraphics{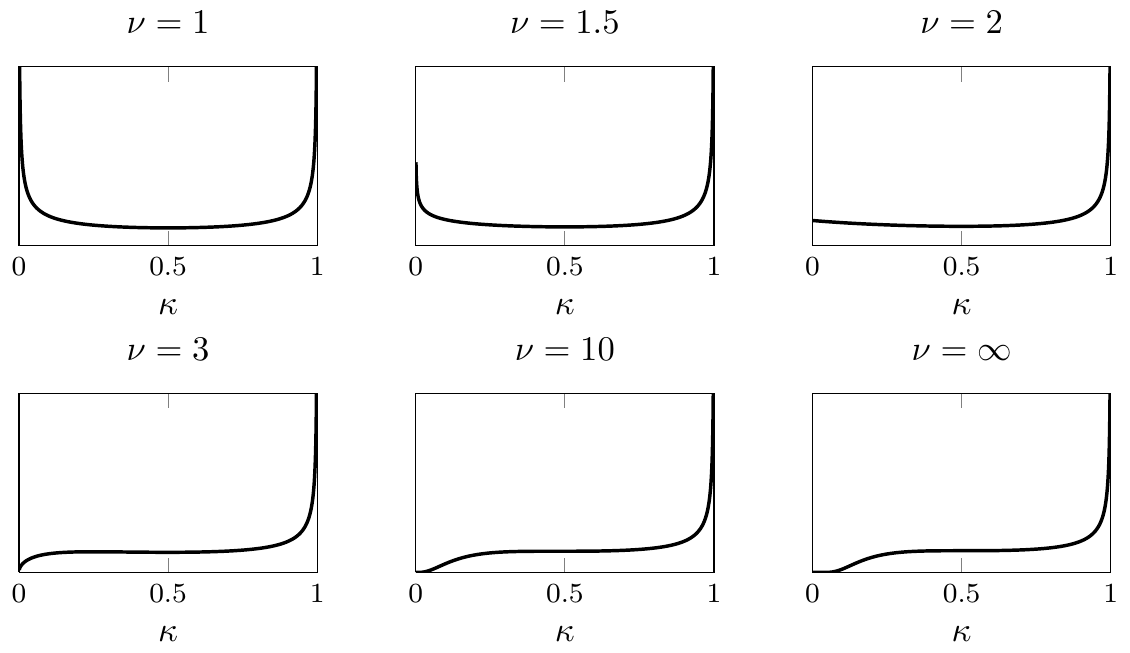}
	\caption{The shrinkage profile of the prior \eqref{eq:hs}, that is, the prior distribution of the shrinkage coefficient $\kappa_i = 1/(1+\lambda_i^2)$, for different values of $\nu$. The case $\nu=1$ corresponds to the horseshoe. $\kappa_i=0$ means no shrinkage and $\kappa_i=1$ complete shrinkage.} 
	\label{fig:shrinkage}
\end{figure}

Figure \ref{fig:shrinkage} shows the priors on $\kappa_i$ implied by different choices of $\nu$.
In all the cases, the prior encourages shrinkage ($\kappa_i\approx 1$) which is due to the fact that the density of the half-$t$ prior evaluates to a positive constant near the origin, and thus contains mass near $\lambda_i=0$.
Moreover, the long tails of the Cauchy distribution ($\nu=1$) allow some of the $\lambda_i$ to be very large, leaving those weights essentially unshrunk ($\kappa_i = 0$).
The reason why the horseshoe yields results closely similar to the spike-and-slab prior \citep{mitchell1988} is due to this dual nature (complete shrinkage or no-shrinkage).
When the value of $\nu$ is increased, the prior still allows strong shrinkage but leaves none of the weights completely unshrunk.

Recently, an extended version of the horseshoe prior, called the horseshoe+ was proposed by \cite{bhadra2015}.
Consider adding another level of local scale parameters to \eqref{eq:hs} as
\begin{align}
	\begin{split}
	w_i \given \lambda_i,\tau &\sim \Normal{0,\lambda_i^2 \eta_i^2 \tau^2} \\
	\lambda_i &\sim \halfStudent[\nu]{0,1}, \\
	\eta_i &\sim \halfStudent[\nu]{0,1}.
\end{split}
\label{eq:hsplus}
\end{align}
We shall refer to the prior \eqref{eq:hsplus} as HS-$t_\nu$+.
The horseshoe+ is then obtained by using half-Cauchy distributions, that is, setting $\nu=1$ (note that the above notation differs slightly from the original paper).
The authors argue that the horseshoe+ obtains an improved behaviour compared to the original horseshoe in ultra-sparse problems, both theoretically and empirically.

The hierarchical priors \eqref{eq:hs} and \eqref{eq:hsplus} are straightforward to implement and use in Stan.
However, in practice we observe that for the particular case of $\nu=1$ (horseshoe and horseshoe+) NUTS produces a lot of divergent transitions even after the warm-up period\footnote{See this thread for outlining the problem \url{https://groups.google.com/d/msg/stan-dev/pt1NNytNVUI/4PHO9ekefBYJ}}.
Experimentally we find that the number of divergent transitions can be reduced by larger value of $\nu$, that is, by shortening the tails of the priors for the local scale terms (see Section~\ref{sec:fitting}).
As depicted in Figure~\ref{fig:shrinkage}, this affects the shrinkage properties of the prior but as will be demonstrated in Section~\ref{sec:fitting}, the effect regarding the predictions may be negligible.
See also the study by \cite{ghosh2015} discussing the problems of the half-Cauchy prior for the logistic regression when the data is separable.

\section{Projection predictive variable selection}
\label{sec:projection}

This section shortly reviews the idea of the projection predictive method for variable selection.
Section~\ref{sec:projection_framework} presents the general idea and Section~\ref{sec:proj_lgm} discusses the linear Gaussian regression model as a special case.
Our discussion is concise, for more information about the assessment of the method, see our previous paper \citep{piironen2015}, and for more on the theoretical considerations and related concepts, see the review by \cite{vehtari2012}.

\subsection{General framework}
\label{sec:projection_framework}

The idea in the projection approach of \cite{goutis1998} and \cite{dupuis2003} is to simplify the full model $M_*$ by projecting the information in the posterior onto the submodels so that the predictions change as little as possible.
Given the parameters of the full model $\vs \theta^*$, the projected parameters $\vs \theta^\perp$ in the parameter space of submodel $M_\perp$ are defined as
\begin{align}
	\vs \theta^\perp 
	&= \arg \min_{\vs \theta} \frac{1}{n}\sum_{i=1}^n
						\KL{p(\ti y \given \vc x_i, \vs \theta^*, M_*)} 
							{p(\ti y \given \vc x_i, \vs \theta, M_\perp)} \,.
\label{eq:projection}
\end{align}
The discrepancy between the full model and the submodel is then defined to be the expectation of this divergence over the posterior of the full model
\begin{align}
	\delta(M_* \| M_\perp)
	= \frac{1}{n}\sum_{i=1}^n
	\mean[ \vs \theta^* \mid \data, M_* ] {
	\KL{p(\ti y \given \vc x_i, \vs \theta^*, M_*)}
		{p(\ti y \given \vc x_i, \vs \theta^\perp, M_\perp)} } \,.
\label{eq:projection_discrepancy}
\end{align}
The posterior expectation in~\eqref{eq:projection_discrepancy} is in general not available analytically.
\cite{dupuis2003} proposed calculating the discrepancy by drawing samples $\{\vs \theta_s^*\}_{s=1}^S$ from the posterior of the reference model, calculating the projected parameters $\{\vs \theta_s^\perp\}_{s=1}^S$ individually according to~\eqref{eq:projection}, and then approximating~\eqref{eq:projection_discrepancy} as
\begin{align}
	\delta(M_* \| M_\perp) \approx
	\frac{1}{n S}\sum_{i=1}^n \sum_{s=1}^S
	\KL{p(\ti y \mid \vc x_i, \vs \theta_s^*, M_*)}
		{p(\ti y \mid \vc x_i, \vs \theta_s^\perp, M_\perp)} \,.
\label{eq:projection_discrepancy_approx}
\end{align}
This approximation makes the computations feasible as it requires only ability to draw samples from the full model and a routine for solving the optimization problem~\eqref{eq:projection}.
In general case the optimization problem can be solved numerically (using e.g. Newton's method) but for the simplest models such as the linear Gaussian case, the minimization can be carried out analytically (see the next section). 

In model selection, we seek for submodels $M_\perp$ which have small discrepancy from the full model~\eqref{eq:projection_discrepancy_approx}.
The final assessment of how much the full model can be simplified, can be performed using cross-validation (see Section~\ref{sec:selection}).

\subsection{Example: linear Gaussian model}
\label{sec:proj_lgm}

Consider the linear Gaussian model \eqref{eq:lgm}.
For easier notation, we set the first term of $\vc w$ to be the intercept $w_0$ and the first input to be fixed $x_0=1$.
For this model, the projected parameters \eqref{eq:projection} can be calculated analytically.
Given a sample $(\vc w,\sigma^2)$ from the posterior of the full model, the projected parameters are given by (see Appendix \ref{app:lgm_projection})
\begin{align}
	\label{eq:proj_weight}
	\vc w_\perp &= ({\vc X_\perp}^\tp \vc X_\perp)^{-1} {\vc X_\perp}^\tp  \vc X \vc w  \\
	\label{eq:proj_noise}
	\sigma_\perp^2 &= \sigma^2 + \frac{1}{n} (\vc X \vc w- \vc X_\perp \vc w_\perp)^\tp (\vc X \vc w- \vc X_\perp \vc w_\perp),
\end{align}
and the associated KL-divergence (for this particular sample) is
\begin{align}
	d(\vc w,\sigma^2) = \frac{1}{2}\log \frac{\sigma^2_\perp}{\sigma^2}.
\label{eq:proj_kl}
\end{align}
Here $\vc X = (\vc x_1^\tp,\dots,\vc x_n^\tp)$ denotes the $n\times m$ predictor matrix of the full model, and $\vc X_\perp$ the contains those columns of $\vc X$ that correspond to the submodel we are projecting onto.
The projection equations \eqref{eq:proj_weight} and \eqref{eq:proj_noise} have a nice interpretation.
The projected weights are determined by the maximum likelihood solution with the observations $\vc y$ replaced by the fit of the full model $\vc f = \vc X \vc w$.
The projected noise variance is the noise level of the full model plus the mismatch between the fits of the full and the projected model.

As discussed in the previous section, we draw a sample $\{\vc w_s,\sigma^2_s\}_{s=1}^S$ from the posterior of the full model, compute the projected parameters and associated KL-divergences according to Equations~\eqref{eq:proj_weight}, \eqref{eq:proj_noise} and \eqref{eq:proj_kl}, and then estimate the discrepancy between the full and submodel as
\begin{align}
	\delta(M_*\|M_\perp) = \frac{1}{S} \sum_{s=1}^S d(\vc w_s,\sigma_s^2)\,.
\label{eq:proj_lgm_discrepancy}
\end{align}
This procedure will produce a parsimonious model with exactly zero weights for the variables that are left-out.

\section{Numerical example: Crime dataset}
\label{sec:numexample}

This section presents an example of fitting a regression model under the HS-$t_\nu$ and HS-$t_\nu$+ priors using Stan (Section~\ref{sec:fitting}), and demonstrates how to perform the subsequent variable selection using R (Section~\ref{sec:selection}).
All the codes can be found online\footnote{\url{https://github.com/jtpi/rstan-varsel}}, and the most relevant parts are also included in the Appendix.
We use the Crime dataset\footnote{\url{https://archive.ics.uci.edu/ml/datasets/Communities+and+Crime}} used in our earlier study.
After removing the features and instances with missing values, the data consist of $1992$ instances with $d=102$ predictor variables.
We normalized all the input variables to have zero mean and unit variance, and log normalized the original target variable (total number of crimes per population) to get a real valued and more Gaussian output.
For illustrational purposes, we split the data randomly into two so that $n=1000$ points are used for model training and variable selection, and the remaining $\ti n=992$ points are used for testing.

\subsection{Model fitting}
\label{sec:fitting}

We fit the linear Gaussian regression model \eqref{eq:lgm} with all the $p=102$ predictors under the HS-$t_\nu$ and HS-$t_\nu$+ priors with $\nu = 1$ and $\nu=3$ (see Section~\ref{sec:hier_shrinkage}).
Note that we use these hierarchical priors only for the weights of the nonconstant inputs.
For the intercept term we use a weakly informative prior
\begin{align*}
	w_0 \sim \Normal{0,5^2}.
\end{align*}
For the global scale parameter $\tau$ and for the noise variance $\sigma^2$, we use the following uninformative priors
\begin{align*}
	\tau &\sim \halfCauchy{0,1} \\
	\sigma^2 &\propto 1 \,.
\end{align*}
The Stan codes for fitting the models are given in Appendix \ref{app:stancodes}.

\begin{figure}
\centering
	\includegraphics{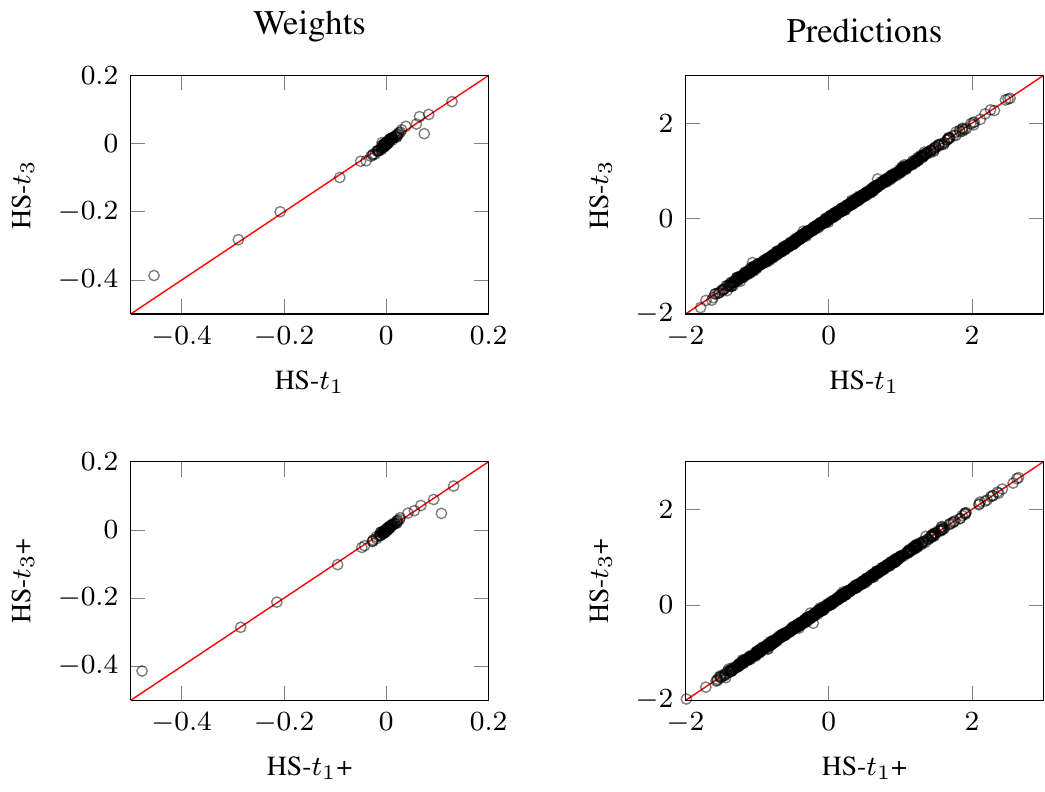}
	\caption{Left column: Posterior means for the weights with different priors, HS-$\StudentNP[1]$ vs. HS-$\StudentNP[3]$ (top) and HS-$\StudentNP[1]$+ vs. HS-$\StudentNP[3]$+ (bottom). HS-$\StudentNP[1]$ and HS-$\StudentNP[1]$+ correspond to horseshoe and horsehoe+ (see Section~\ref{sec:hier_shrinkage}). Right column: the same but for the predictive means on the test set.}
	\label{fig:hs_vs_t}
\end{figure}

Using the training data, we sample 4 chains each having 1000 samples, and from each chain we remove the first half as a warmup.
Figure~\ref{fig:hs_vs_t} shows the effect of the degrees of freedom $\nu$ in \eqref{eq:hs} and \eqref{eq:hsplus} on the posterior means of the weights and predictions on the test data.
In both cases, two of the weights are shrunk more heavily towards zero under $\nu=3$, while the predictions remain practically the same.
The percentage of divergent transitions under the posterior simulations were $3.4\%$ and $5.2\%$ for HS-$\StudentNP[1]$ and HS-$\StudentNP[1]$+, whereas only $0.0\%$ and $0.1\%$ for HS-$\StudentNP[3]$ and HS-$\StudentNP[3]$+.
Given the more robust sampling and negligible effect on the predictive performance, we proceed to the variable selection using $\nu=3$.
However, we emphasize the tentative nature of this experiment and do not recommend this choice to be used uncritically.
Thorough understanding of the effects of $\nu$ needs more research, but we do not focus on it in this report.

\subsection{Variable selection}
\label{sec:selection}

After fitting the model (previous section) with all the variables we proceed to the variable selection.
Here we use the HS-$\StudentNP[3]$ prior (Section~\ref{sec:hier_shrinkage}) for the full model and the projection predictive variable selection strategy (Section~\ref{sec:proj_lgm}).
As a search heuristic, we use forward searching, that is, starting from the empty model, we add variables one at a time, each time choosing the variable that decreases the KL-divergence \eqref{eq:proj_lgm_discrepancy} the most.
As discussed in our study \citep{piironen2015}, the effect of number of chosen variables on the predictive ability can be assessed reliably using cross-validation.
In other words, we repeat the model fitting and selection $K$ times each time leaving $n/K$ points out for validation, and evaluate the performance of the full model and the found submodels using these left-out data.
The relevant R codes are provided in Appendix~\ref{app:Rcodes}.

\begin{figure}
\centering
	\includegraphics{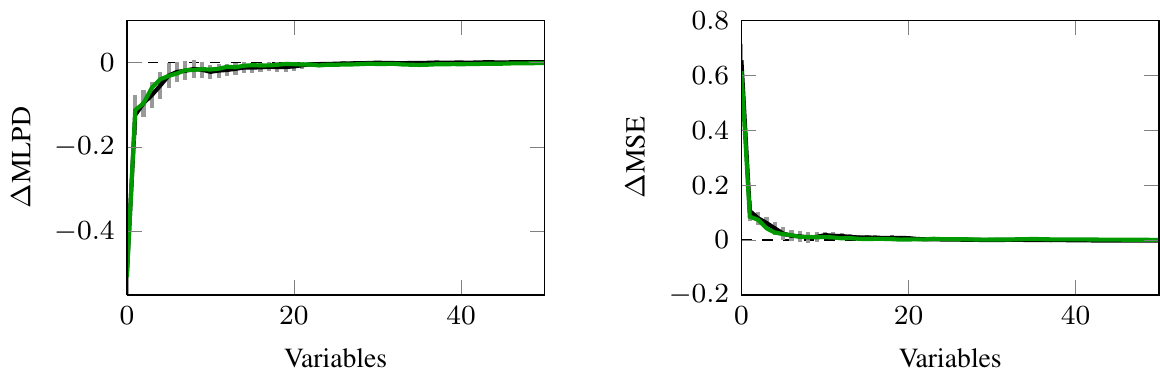}
	\caption{Difference in the mean log predictive density (MLPD) and mean squared error (MSE) between the submodel and the full model as a function of number of chosen variables up to 50 variables. Black is the average over the $K=10$ cross-validated searches, grey bars denoting the 95\% credible interval, and green is the test performance when the search is done using all the training data.}
	\label{fig:lpd_mse}
\end{figure}

Figure \ref{fig:lpd_mse} shows the difference in the mean log predictive density and mean squared error between the submodel and the full model as a function of number of added variables up to 50 variables.
The black line is the average over the $K=10$ cross-validation folds and the green line shows the result when the fitting and searching is performed using all the training data and the performance is evaluated on the test data.
As expected, there is a good correspondence between the cross-validated and test performance.
For this dataset, most of the predictive ability of the full model is captured with about 5 variables, and 20 variables are enough for getting predictions indistinguishable from the full model for all practical purposes.

\bibliographystyle{apalike}
\bibliography{references}

\begin{thebibliography}{}

\bibitem[Bhadra et~al., 2015]{bhadra2015}
Bhadra, A., Datta, J., Polson, N.~G., and Willard, B. (2015).
\newblock The horseshoe+ estimator of ultra-sparse signals.
\newblock {\em arXiv:1502.00560}.

\bibitem[Carvalho et~al., 2009]{carvalho2009}
Carvalho, C.~M., Polson, N.~G., and Scott, J.~G. (2009).
\newblock Handling sparsity via the horseshoe.
\newblock In {\em Proceedings of the 12th International Conference on
  Artificial Intelligence and Statistics}, pages 73--80.

\bibitem[Carvalho et~al., 2010]{carvalho2010}
Carvalho, C.~M., Polson, N.~G., and Scott, J.~G. (2010).
\newblock The horseshoe estimator for sparse signals.
\newblock {\em Biometrika}, 97(2):465--480.

\bibitem[Datta and Ghosh, 2013]{datta2013}
Datta, J. and Ghosh, J.~K. (2013).
\newblock Asymptotic properties of {B}ayes risk for the horseshoe prior.
\newblock {\em {Bayesian Analysis}}, 8(1):111--132.

\bibitem[Dupuis and Robert, 2003]{dupuis2003}
Dupuis, J.~A. and Robert, C.~P. (2003).
\newblock Variable selection in qualitative models via an entropic explanatory
  power.
\newblock {\em Journal of Statistical Planning and Inference}, 111(1-2):77--94.

\bibitem[Ghosh et~al., 2015]{ghosh2015}
Ghosh, J., Li, Y., and Mitra, R. (2015).
\newblock On the use of {C}auchy prior distributions for {B}ayesian logistic
  regression.
\newblock {\em arXiv:1507.07170}.

\bibitem[Goutis and Robert, 1998]{goutis1998}
Goutis, C. and Robert, C.~P. (1998).
\newblock Model choice in generalised linear models: a {B}ayesian approach via
  {K}ullback--{L}eibler projections.
\newblock {\em Biometrika}, 85(1):29--37.

\bibitem[Mitchell and Beauchamp, 1988]{mitchell1988}
Mitchell, T.~J. and Beauchamp, J.~J. (1988).
\newblock {B}ayesian variable selection in linear regression.
\newblock {\em {Journal of the American Statistical Association}},
  83(404):1023--1036.

\bibitem[Piironen and Vehtari, 2015]{piironen2015}
Piironen, J. and Vehtari, A. (2015).
\newblock Comparison of {B}ayesian predictive methods for model selection.
\newblock {\em arXiv:1503.08650}.

\bibitem[van~der Pas et~al., 2014]{varDerPas2014}
van~der Pas, S.~L., Kleijn, B. J.~K., and van~der Vaart, A.~W. (2014).
\newblock The horseshoe estimator: posterior concentration around nearly black
  vectors.
\newblock {\em {Electronic Journal of Statistics}}, 8(2):2585--2618.

\bibitem[Vehtari and Ojanen, 2012]{vehtari2012}
Vehtari, A. and Ojanen, J. (2012).
\newblock A survey of {B}ayesian predictive methods for model assessment,
  selection and comparison.
\newblock {\em Statistics Surveys}, 6:142--228.

\end{thebibliography}

\section*{Acknowledgments}

We thank Tomi Peltola for sharing his Stan-codes which helped implementing the models in this document.

\setcounter{equation}{0}

\begin{appendices}

\renewcommand{\theequation}{\Alph{section}.\arabic{equation}}

\section{Projection for linear Gaussian model}
\label{app:lgm_projection}

Let $f_i^\perp =  {\vc w_\perp}^\tp \vc x_i^\perp$ denote the fit of the submodel, where $\vc w_\perp$ is the projected weight vector and $\vc x_i^\perp$ the input vector corresponding to the submodel on which we are projecting.
Moreover, let $\sigma^2_\perp$ denote the projected noise variance.
The cost function in \eqref{eq:projection} can be written as
\begin{align}
	&\frac{1}{n}\sum_{i=1}^n \KL{ \Normal{f_i ,\sigma^2}  }
						{ \Normal{f_i^\perp ,\sigma_\perp^2} }  \nonumber \\ 
	&= \frac{1}{n}\sum_{i=1}^n 
		\frac{1}{2}\left(\log \frac{\sigma_\perp^2}{\sigma^2} + \frac{\sigma^2 + (f_i - f_i^\perp)^2}{\sigma_\perp^2} -1 \right) \nonumber \\
	&= \frac{1}{2}\left(\log \frac{\sigma_\perp^2}{\sigma^2} + \frac{1}{\sigma_\perp^2} \left(\sigma^2 +  \frac{1}{n}\sum_{i=1}^n (f_i - f_i^\perp)^2 \right)  - 1 \right) \nonumber \\
	&= \frac{1}{2}\left(\log \frac{\sigma_\perp^2}{\sigma^2} + \frac{1}{\sigma_\perp^2} \left(\sigma^2 +  \frac{1}{n} (\vc f- \vc f_\perp)^\tp (\vc f- \vc f_\perp) \right)  -1 \right), \nonumber \\
	&= \frac{1}{2}\left(\log \frac{\sigma_\perp^2}{\sigma^2} + \frac{1}{\sigma_\perp^2} \left(\sigma^2 +  \frac{1}{n} (\vc X \vc w- \vc X_\perp \vc w_\perp)^\tp (\vc X \vc w- \vc X_\perp \vc w_\perp) \right) -1 \right),
	\label{aeq:projkl}
\end{align}
where $\vc f=(f_1,\dots,f_n)$, $\vc f_\perp=(f_1^\perp,\dots,f_n^\perp)$, and $\vc X$ and $\vc X_\perp$ denote the predictor matrices of the full and submodel, respectively.
Setting the gradient of \eqref{aeq:projkl} with respect to $\vc w_\perp$ to zero, we obtain
\begin{align}
	\vc w_\perp = ({\vc X_\perp}^\tp \vc X_\perp)^{-1} {\vc X_\perp}^\tp  \vc X \vc w. %
\label{aeq:proj_weight}
\end{align}
Accordingly, minimizing \eqref{aeq:projkl} with respect to $\sigma_\perp^2$  gives us
\begin{align}
	\sigma_\perp^2 = \sigma^2 + \frac{1}{n} (\vc X \vc w- \vc X_\perp \vc w_\perp)^\tp (\vc X \vc w- \vc X_\perp \vc w_\perp).
\label{aeq:proj_noise}
\end{align}
Plugging the expressions \eqref{aeq:proj_weight} and \eqref{aeq:proj_noise} into \eqref{aeq:projkl} gives us the divergence introduced by the projection of these particular parameters $(\vc w,\sigma^2)$
\begin{align}
	d(\vc w,\sigma^2) &= \frac{1}{n}\sum_{i=1}^n \KL{ \Normal{f_i ,\sigma^2}  }
						{ \Normal{f_i^\perp ,\sigma_\perp^2} } \nonumber \\
					&= \frac{1}{2} \log \frac{\sigma^2}{\sigma_\perp^2}.
\end{align}

\section{Stan codes}
\label{app:stancodes}
\lstset{keywordstyle=\color{black}}

Linear regression with Gaussian noise and HS-$\StudentNP[\nu]$ prior \eqref{eq:hs} on the weights. The case $\nu=1$ corresponds to the horseshoe.
\begin{lstlisting}[language=C, tabsize=4, basicstyle=\ttfamily\footnotesize]

/* lg_t.stan */

functions {
	// square root of a vector (elementwise)
	vector sqrt_vec(vector x) {
		vector[dims(x)[1]] res;

		for (m in 1:dims(x)[1]){
			res[m] <- sqrt(x[m]);
		}
		return res;
	}
}

data {
	int<lower=0> n; // number of observations
	int<lower=0> d; // number of predictors
	vector[n] y;	// outputs
	matrix[n,d] X;	// inputs
	real<lower=1> nu; // degrees of freedom for the half t-priors
}

parameters {

	// intercept and noise std
	real w0;
	real<lower=0> sigma;
  
	// auxiliary variables for the variance parameters
	vector[d] z;
	real<lower=0> r1_global;
	real<lower=0> r2_global;
	vector<lower=0>[d] r1_local;
	vector<lower=0>[d] r2_local;
}

transformed parameters {
	
	// global and local variance parameters, and the input weights
	real<lower=0> tau;
	vector<lower=0>[d] lambda;
	vector[d] w;
	
	tau <- r1_global * sqrt(r2_global);
	lambda <- r1_local .* sqrt_vec(r2_local);
	w <- z .* lambda*tau;
}

model {
	
	// observation model
	y ~ normal(w0 + X*w, sigma);
	
	// half t-priors for lambdas (nu = 1 corresponds to horseshoe)
	z ~ normal(0, 1);
	r1_local ~ normal(0.0, 1.0);
	r2_local ~ inv_gamma(0.5*nu, 0.5*nu);
	
	// half cauchy for tau
	r1_global ~ normal(0.0, 1.0);
	r2_global ~ inv_gamma(0.5, 0.5);
	
	// weakly informative prior for the intercept
	w0 ~ normal(0,5);
	
	// using uniform prior on the noise variance
}
\end{lstlisting}

\vspace{0.5cm}
\noindent The same model as above, but with HS-$\StudentNP[\nu]$+ prior \eqref{eq:hsplus} on the weights. The case $\nu=1$ corresponds to the horseshoe+.
\begin{lstlisting}[language=C, tabsize=4, basicstyle=\ttfamily\footnotesize]

/* lg_tplus.stan */

functions {
	// square root of a vector (elementwise)
	vector sqrt_vec(vector x) {
		vector[dims(x)[1]] res;

		for (m in 1:dims(x)[1]){
			res[m] <- sqrt(x[m]);
		}
		return res;
	}
}

data {
	int<lower=0> n; // number of observations
	int<lower=0> d; // number of predictors
	vector[n] y;	// outputs
	matrix[n,d] X;	// inputs
	real<lower=1> nu; // degrees of freedom for the half t-priors
}

parameters {

	// intercept and noise std
	real w0;
	real<lower=0> sigma;
  
	// auxiliary variables for the variance parameters
	vector[d] z;
	real<lower=0> r1_global;
	real<lower=0> r2_global;
	vector<lower=0>[d] r1_local;
	vector<lower=0>[d] r2_local;
	vector<lower=0>[d] r1_localplus;
	vector<lower=0>[d] r2_localplus;
}

transformed parameters {
	
	// global and local variance parameters, and the input weights
	real<lower=0> tau;
	vector<lower=0>[d] lambda;
	vector<lower=0>[d] lambdaplus;
	vector[d] w;
	
	tau <- r1_global * sqrt(r2_global);
	lambda <- r1_local .* sqrt_vec(r2_local);
	lambdaplus <- r1_localplus .* sqrt_vec(r2_localplus);
	w <- z .* lambda .* lambdaplus * tau;
}

model {
	
	// observation model
	y ~ normal(w0 + X*w, sigma);
	
	// half t-priors for all the lambdas (nu = 1 corresponds to horseshoe+)
	z ~ normal(0, 1);
	r1_local ~ normal(0.0, 1.0);
	r2_local ~ inv_gamma(0.5*nu, 0.5*nu);
	r1_localplus ~ normal(0.0, 1.0);
	r2_localplus ~ inv_gamma(0.5*nu, 0.5*nu);
	
	// half cauchy for tau
	r1_global ~ normal(0.0, 1.0);
	r2_global ~ inv_gamma(0.5, 0.5);
	
	// weakly informative prior for the intercept
	w0 ~ normal(0,5);
	
	// using uniform prior on the noise variance
}
\end{lstlisting}

\section{R codes}
\label{app:Rcodes}

Code for performing the projection onto a submodel for linear Gaussian model, given a sample from the full model:
\begin{lstlisting}[language=R, tabsize=4, basicstyle=\ttfamily\footnotesize]
lm_proj <- function(w,sigma2,x,indproj) {
	
	# assume the intercept term is stacked into w, and x contains 
	# a corresponding vector of ones. returns the projected samples
	# and estimated kl-divergence.

	# pick the columns of x that form the projection subspace
	n <- dim(x)[1]
	xp <- x[,indproj]

	# solve the projection equations
	fit <- x %*% w # fit of the full model
	wp <- solve(t(xp) %*% xp, t(xp) %*% fit)
	sigma2p <- sigma2 + colMeans((fit - xp %*% wp)^2)
	
	# this is the estimated kl-divergence between the full and projected model
	kl <- mean(0.5*log(sigma2p/sigma2))
	
	# reshape wp so that it has same dimensionality as x, and zeros for
	# those variables that are not included in the projected model
	d <- dim(w)[1]
	S <- dim(w)[2]
	wptemp <- matrix(0, d, S)
	wptemp[indproj,] <- wp
	wp <- wptemp
	
	return(list(w=wp, sigma2=sigma2p, kl=kl))
}
\end{lstlisting}

\vspace{0.5cm}
\noindent Forward variable searching by minimizing the associated KL-divergence in the projection:
\begin{lstlisting}[language=R, tabsize=4, basicstyle=\ttfamily\footnotesize]
lm_fprojsel <- function(w, sigma2, x) {

	# forward variable selection using the projection. returns the
	# indices of the variables chosen (in order) and accociated kl-divergences.
	
	d = dim(x)[2]
	chosen <- 1 # chosen variables, start from the model with the intercept only
	notchosen <- setdiff(1:d, chosen)
	
	# start from the model having only the intercept term
	kl <- rep(0,d)
	kl[1] <- lm_proj(w,sigma2,x,1)$kl

	# start adding variables one at a time
	for (k in 2:d) {
	
		nleft <- length(notchosen)
		val <- rep(0, nleft)

		for (i in 1:nleft) {
			ind <- sort( c(chosen, notchosen[i]) )
			proj <- lm_proj(w,sigma2,x,ind)
			val[i] <- proj$kl
		}

		# find the variable that minimizes the kl
		imin <- which.min(val)
		chosen <- c(chosen, notchosen[imin])
		notchosen <- setdiff(1:d, chosen)
	
		kl[k] <- val[imin]
	}
	return(list(chosen=chosen, kl=kl))
}
\end{lstlisting}

\clearpage
\vspace{0.5cm}
\noindent Example of sampling the full model and performing the variable selection using the projection:
\begin{lstlisting}[language=R, tabsize=4, basicstyle=\ttfamily\footnotesize]
# load the data
n <- 1000 # number of training points
data <- as.matrix(read.table('crimedata.csv', sep=','))
dims <- dim(data)
ntotal <- dims[1]
nt <- ntotal - n # number of test points
d <- dims[2]-1

# training data
y <- data[1:n, 1]
x <- data[1:n, 2:(d+1)]

# test data
yt <- data[(n+1):ntotal, 1]
xt <- data[(n+1):ntotal, 2:(d+1)]

# fit the full model
fit <- stan("lg_t.stan", data=list(X=x,y=y,d=d,n=n,nu=3.0), iter=100, chains=1)
e <- extract(fit)

w <- rbind(e$w0, t(e$w)) # stack the intercept and weights
sigma2 <- e$sigma^2
x <- cbind(rep(1,n), x) # add a vector of ones to the predictor matrix
xt <- cbind(rep(1,ntest), xt) 

# perform the variable selection
spath <- lm_fprojsel(w, sigma2, x)

# compute the predictions on the test set using the projected parameters
mlpd <- rep(0, d+1)
mse <- rep(0, d+1)
for (k in 1:(d+1)) {

	# projected parameters
	submodel <- lm_proj(w, sigma2, x, spath$chosen[1:k])
	wp <- submodel$w
	sigma2p <- submodel$sigma2
		
	# mean squared error
	ypred <- rowMeans(xt %*% wp)
	mse[k] <- mean((yt-ypred)^2)
	
	# mean log predictive density
	pd <- dnorm(yt, xt %*% wp, sqrt(sigma2p))
	mlpd[k] <- mean(log(rowMeans(pd)))
}
\end{lstlisting}

\end{appendices}

\end{document}